\documentclass{svjour2}
\title{Constraining the parameters of binary systems through time-dependent
light deflection}
\author{Edmund Schluessel}
\institute{School of Physics and Astronomy, 5, The Parade, Cardiff University,
Cardiff, Wales, United Kingdom CF24 3AA. \\Tel.: +44(0)2920 875107, \\e-mail: Edmund.Schluessel@astro.cf.ac.uk}

\usepackage{times}
\usepackage[T1]{fontenc}
\usepackage[latin1]{inputenc}
\usepackage{graphicx}

\makeatletter

\providecommand{\LyX}{L\kern-.1667em\lower.25em\hbox{Y}\kern-.125emX\@}
\newcommand{\noun}[1]{\textsc{#1}}

\makeatother
\begin{document}

\maketitle

\begin{abstract}
A theory is derived relating the configuration of the cores of active
galaxies, specifically candidates for presumed super-massive black
hole binaries (SMBHBs), to time-dependent changes in images of those
galaxies. Three deflection quantities, resulting from the monopole
term, mass quadrupole term, and spin dipole term of the core, are examined. The
resulting observational technique is applied to the galaxy 3C66B.
This technique is found to under idealized circumstances surpass the
technique proposed by Jenet \emph{et al.} in accuracy for constraining
the mass of SMBHB candidates, but is exceeded in accuracy and precision
by Jenet's technique under currently-understood likely conditions.
The technique can also under favorable circumstances produce results
measurable by currently-available astronomical interferometry such
as very-long baseline-interferometry (VLBI).
\end{abstract}


\section{Introduction}

The deflection of light by gravity is the oldest experimentally-verified
test of the theory of general relativity \cite{key-5}. With the continued
improvement in observational resolution in astronomy, particularly
through very-long-baseline interferometry (VLBI), the detection of
more subtle effects of this light deflection becomes practical. Consequently,
light deflection can be used to measure the properties of distant
systems. This paper supplies a theory for using time-variable light
deflection to measure or constrain the parameters of binary systems.
Specifically, the deflection angle of a light ray from a distant source
is related to the configuration and motion of a binary system located
in a distant galaxy somewhere between the point of emission of the
light ray and its observation.

Super-massive black hole binaries (SMBHBs) are thought to form the
cores and primary energy sources of the broad class of galaxies termed
{}``active galaxies'', {}``blazars'', or {}``quasars''. However,
a combination of distance, radio noise, and optical thickness makes
direct observation of presumed SMBHBs impractical. Observing a time-dependent
motion in the image of the galaxy can provide information on the mass
and orbital parameters of an SMBHB candidate.

Work by Damour and Esposito-Farese \cite{key-15} and by Kopeikin \emph{et
al.} \cite{key-14} establishes a theory of time-dependent light deflection
by describing the time-\\dependent part of the deflection through the
quadrupole term, which is the lowest-order term resulting from the
mass distrubtion whose effects are practical to evaluate using current
astronomical observational techniques. The work of Mashhoon and Kopeikin \cite{key-41}
in examining gravitomagnetic effects furthermore provides a theory
for evaluating the contribution of the spin dipole of such systems
and complements the work of Einstein \cite{key-43} in providing a
complete theory for stating the location of the deflected image, in
the weak field limit. We generalize these theories to a stronger-field
regime and put constraints on the theory's applicability in this regime.

As a case study of an active galaxy, the theory is applied to the
galaxy 3C66B, a nearby active galaxy with a candidate SMBHB core \cite{key-24},
and theoretical constraints on 3C66B's parameters from a light deflection
experiment are compared to the constraints claimed by Jenet \emph{et
al}. \cite{key-8}.

\section{Theory}

\subsection{Notations, definitions \& assumptions}

We assume that Einstein's theory of general relativity is true to
the limits of our ability to observe and applicable to the systems
under examination. We do not address MOND or other post-Einsteinian
models.

Throughout this paper, {}``emitter'' refers to the source of light
rays being observed; {}``deflector'' refers to the mass distribution
causing a change in the metric of spacetime from flatness; and {}``observer''
refers to the point where the light rays produced by the emitter are
observed.

We also make use of a coordinate system derived from the Cartesian
system, defined thus: in a space that is asymptotically Cartesian
let a line be described by%
\footnote{Throughout this paper, Greek indices indicate $\left(0,1,2,3\right)$
and Latin indices indicate $\left(1,2,3\right)$. A contravariant
3-vector is denoted either in boldface or with a raised Latin index.%
} \begin{equation}
x^{i}\left(t\right)=k^{i}\left(t-t_{0}\right)+x_{0}^{i} \, .\label{eq:xoft}
\end{equation}
Let $t^{*}$be the time associated with the line's closest approach
to the origin of the Cartesian system. Let $\tau=t-t^{*}$ denote
a new time coordinate (that is, at $\tau=0$ the line reaches the
closest point to the origin of both the Cartesian and projected systems).
Space coordinates are projected onto a plane passing through the origin
of the coordinate system and perpendicular to a line from the observer
to the origin of the coordinate system; these new space coordinates
are denoted $\xi^{i}=P^{ij}x^{j}\left(t^{*}\right)$ where the projection
operator is defined $P^{ij}\equiv\delta^{ij}-k^{i}k^{j}$. In the projected
coordinate system, the index 0 refers to $\tau$ and the indices $i$
denote coordinates $\xi^{i}$.

For a trajectory described by (\ref{eq:xoft}) let $\xi^{j}\equiv P^{ij}\left.x^{i}\left(\tau\right)\right|_{\tau=0}$
be the {}``vector impact parameter'' of the trajectory and let $d\equiv\left|\xi^{i}\right|$
be the {}``scalar impact parameter'' of the trajectory. Since the
space is asymptotically flat, \emph{d} is also the ratio of the magnitudes
of the angular and linear momenta of the light ray. Note then that
for the trajectory described by $x^{i}\left(\tau\right)$, $r\left(\tau\right)=\sqrt{d^{2}+\tau^{2}}$%
\footnote{Throughout this paper we use the convention $G=c=1$ to simplify our
equations.%
}. Let the unit vector $n^{i}\equiv\xi^{i}/d$.

We assume that the wavelengths of all light rays observed are much
shorter than the longest wavelength of gravitational radiation emitted
by the deflecting system.

\subsection{General theory}

\subsubsection{Background}

Consider a photon emitted at some distant point $x_{0}^{i}$ at some
time in the distant past $t_{0}$. This beam of light in asymptotically
flat space follows a path $k^{i}$ such that the coordinate $x^{i}$
of the photon is given by the relation $(\ref{eq:xoft})$; therefore,
$k^{i}=\left.\left(\partial x^{i}/\partial t\right)\right|_{t=-\infty}$.
Let $k^{i}$ be normalized such that $k^{i}k_{i}=1$; then the vector
$k^{\alpha}=\left(1,k^{i}\right)$ is parallel to the four-momentum
of the photon in flat space.

Let an asymptotically-flat metric $g_{\alpha\beta}$%
\footnote{All metrics $g_{\mu\nu}$ in this paper are stated using the harmonic
gauge condition, that is, $g^{\mu\nu}\Gamma_{\mu\nu}^{\lambda}=0$.
The Minkowski metric in Cartesian coordinates is chosen with signature
$\left(-,+,+,+\right)$ and is denoted $\eta_{\mu\nu}$, and we make
use of the Einstein summation convention.%
} be a function of some affine parameter $\lambda$. Let $K^{\alpha}\equiv k^{\alpha}+\kappa^{\alpha}\left(\lambda\right)+\Xi^{\alpha}\left(\lambda\right)$
be the trajectory of a photon moving in this metric space, where $\kappa^{\alpha}$
describes the part of the trajectory arising from the spherically-symmetric
non-flat part of the metric and $\Xi^{\alpha}$ describes the trajectory
arising from a perturbation to the metric. Then, we have the geodesic
equation \cite[equation 87.3]{key-11} 

\begin{equation}
\frac{d\left(\kappa^{\alpha}+\Xi^{\alpha}\right)}{d\lambda}+\Gamma_{\beta\gamma}^{\alpha}K^{\beta}K^{\gamma}=0.\label{geodesic}\end{equation}
The quantity $d\left(\kappa^{\alpha}+\Xi^{\alpha}\right)/d\lambda$
corresponds to the change in momentum of the light ray in space, which
when projected onto a plane of observation corresponds to the angular
deflection of the light ray. We define this deflection vector by \cite{key-14,key-15}
\begin{equation}
\alpha^{i}\left(t,\xi^{i}\right)\equiv P_{j}^{i}\left[\kappa^{j}+\Xi^{j}\right]_{\textrm{observer}}+\Delta\alpha^{i}\label{alpha},\end{equation}
where the term $\Delta\alpha^{i}$ corresponds to corrections arising
from any contribution to deflection other than our deflector.

In the case of Eddington's experiment \cite{key-5} on solar deflection,
the {}``true'' position of the emitter -- that is, the position
of the emitter observed in the limit of intervening deflection going
to zero -- was known. In the case of deflectors with small proper
motion, in this case extragalactic or otherwise distant objects, where
the emitter would be seen without the presence of the intervening deflector may not
be known; therefore, the periastron of the light ray must be determined
by other means. Let \emph{P} be the periastron of the light ray's
trajectory about the deflector; in such cases, the time delay between
the deflection and the motion of the deflector is related to the periastron
by
\begin{equation}
P=t_{\textrm{peak deflection}}-t_{\textrm{alignment}},\label{eq:impactconstraint}
\end{equation}
where $t_{\textrm{peak deflection}}$ is the time when the image of
the source is observed to be deflected most from the position of the
deflector and $t_{\textrm{alignment}}$ is the time when the projected images of the components
of the system and the image fall into a line, assuming that $P<p/2$
and that the change in the gravitational field propagates at the speed
of light.

\subsubsection{Description of deflector}

\begin{figure}[p]
\includegraphics[width=1\textwidth]{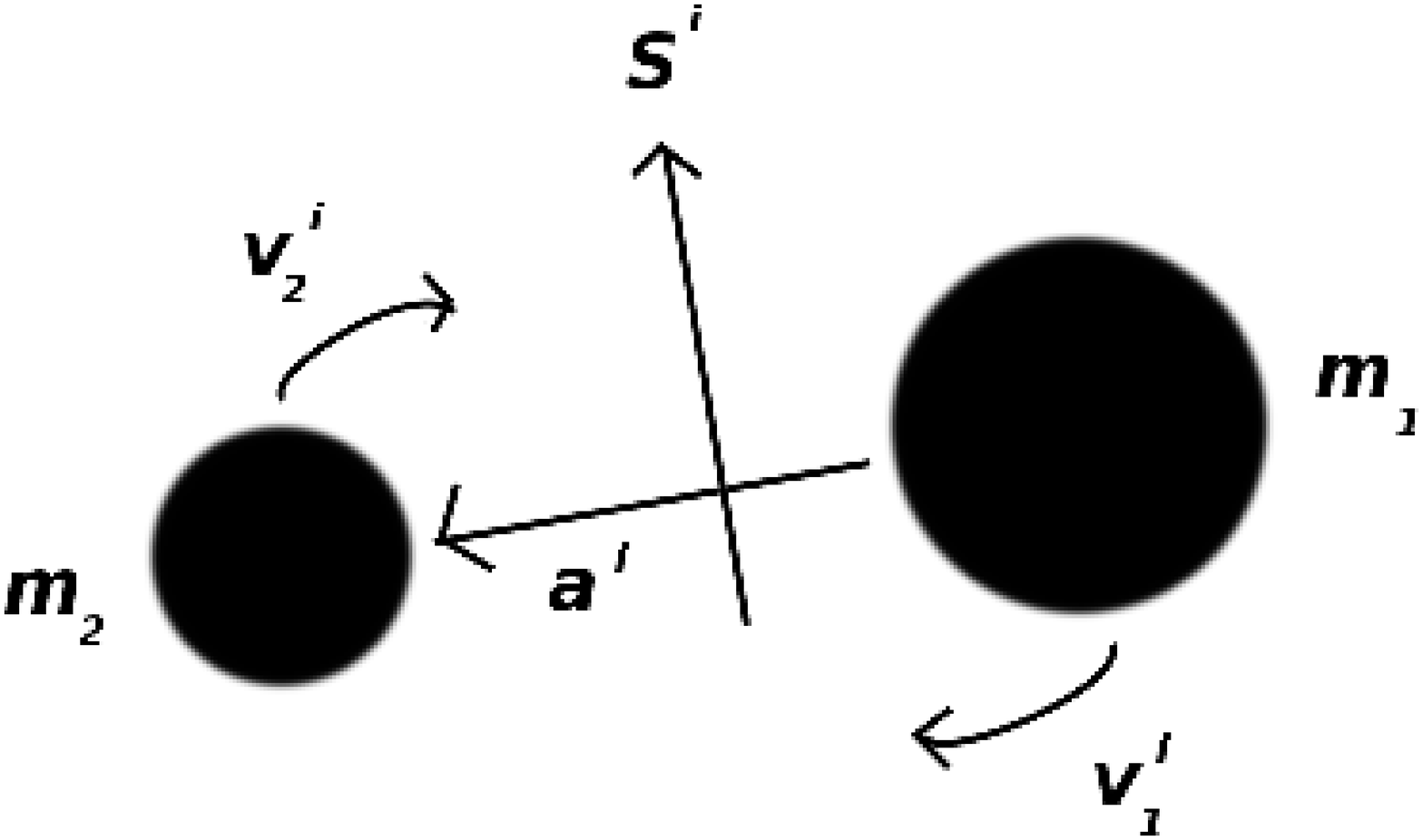}
\caption{\label{cap:Figure 1}Object 1 has mass $m_{1}$, velocity $v_{1}^{i}$
and is located as position $x_{1}^{i}$; object 2 has corresponding
$m_{2}$, $v_{2}^{i}$ and $x_{2}^{i}$. $x_{1}^{i}-x_{2}^{i}=a^{i}$
and the spin vector $S^{i}$ where $S^{i}\equiv\left(J^{23},J^{31},J^{12}\right)$
and $J^{ij}\equiv\int\left(x^{i}T^{j0}-x^{j}T^{i0}\right)dV$ is perpendicular
to $a^{i}$, $v_{1}^{i}$ and $v_{2}^{i}$.}
\end{figure}

Our deflector of interest is as follows: two objects are denoted with
the indices 1 and 2. The mass of object 1 $m_{1}\geq m_{2}$. The
objects have positions $x_{1}^{i}\left(t\right)$ and $x_{2}^{i}\left(t\right)$
and velocities $v_{1}^{i}\left(t\right)$ and $v_{1}^{i}\left(t\right)$\noun{{[}Figure}
\ref{cap:Figure 1}{]}. Then our source has density distribution\begin{equation}
\rho\left(t,x^{i}\right)=m_{1}\delta\left(x^{i}-x_{1}^{i}\left(t\right)\right)+m_{2}\delta\left(x^{i}-x_{2}^{i}\left(t\right)\right)\label{density}\end{equation}
and velocity distribution
\begin{eqnarray}
v^{i}\left(t,x^{j}\right) & = & v_{1}^{i}\delta\left(x^{j}-x_{1}^{j}\right)+v_{2}^{i}\delta\left(x^{j}-x_{2}^{j}\right)\label{velocity}
\end{eqnarray}
where $\delta\left(x^{i}\right)$ is the three-dimensional Dirac delta
distribution.

Our metric has the form $g_{\mu\nu}=\eta_{\mu\nu}+s_{\mu\nu}+h_{\mu\nu}$
where $s_{\mu\nu}$ is the non-Minkowski part of the Schwarzschild metric and $h_{\mu\nu}$
is a small perturbation. Let $h_{\mu\nu}^{Q}$ be the perturbation
resulting from the quadrupole moment of the mass distribution and
let $h_{\mu\nu}^{S}$ be the perturbation resulting from the spin
dipole of the mass distrubtion. Let the variable $s=t-r$. Then explicitly,
the metric is given by \cite[p 181]{key-15,key-45,key-41,key-42}:
\begin{eqnarray}
s_{00} & = & \frac{2m}{r}, \nonumber \\
s_{0i} & = & 0, \label{eq:smetric}\\
s_{ij} & = & \left[\left(1-\frac{2m}{r}\right)^{-1}-1\right]\frac{x^{i}}{r^{2}}\delta_{ij};  \nonumber
\end{eqnarray}
and
\begin{eqnarray}
h_{00}^{Q} & = & \frac{\partial^{2}}{\partial x^{i}\partial x^{j}}\frac{Q_{ij}\left(s\right)}{r}, \nonumber\\
h_{0i}^{Q} & = & -2\frac{\partial^{2}}{\partial x^{i}\partial t}\frac{Q_{ij}\left(s\right)}{r}, \label{eq:hmetric} \\
h_{ij}^{Q} & = & \frac{\partial^{2}}{\partial x^{i}\partial x^{j}}\frac{Q_{ij}\left(s\right)}{r}\delta_{ij}+2\frac{\partial^{2}}{\partial t^{2}}\frac{Q_{ij}\left(s\right)}{r}; \nonumber
\end{eqnarray}
and
\begin{eqnarray}
h_{0i}^{S} & = & 2\frac{S_{j}x_{k}\epsilon^{i}{}_{jk}}{r^{3}}, \nonumber \\
h_{00}^{S} & = & h_{ij}^{S}=0 \label{eq:hqmetric},
\end{eqnarray}
where the vector $S^{i}\equiv\left(J^{23},J^{31},J^{12}\right)$ and
$J^{ij}\equiv\int\left(x^{i}T^{j0}-x^{j}T^{i0}\right)dV$ \cite[chapter 2.9]{key-42}. 

Let the objects orbit one another with a known period $p$. Let our
coordinate system origin be located at the center of mass of the binary
and let $m=m_{1}+m_{2}$. Let $a^{i}\equiv x_{1}^{i}-x_{2}^{i}$ be
a vector denoting the spatial separation of the two masses and $l\equiv\left|a^{i}\right|$.
Let the mass ratio $q\equiv m_{2}/m_{1}\leq1$.

By our choice of coordinates, the dipole term of the deflector's mass
distribution is zero.

Using the Landau-Lifschitz definition of the transverse traceless\\
quadrupole \cite[eq 41.3]{key-11}, the quadrupole moment of
the deflector is given by:
\begin{equation}
Q_{ij}\left(t\right)=\int\rho\left(\mathbf{x},t\right)\left[3x_{i}x_{j}-r^{2}\delta_{ij}\right]dV=\frac{mq}{\left(1+q\right)^{2}}\left[3a_{i}a_{j}-l^{2}\delta_{ij}\right].\label{quadrupole}
\end{equation}

In the case that the masses travel in almost circular orbits about
their common center of mass, then in a primed coordinate system related
to our chosen system only by unitary rotations, 
\begin{equation}
a^{\prime i}\left(t\right)=l\left(\begin{array}{c}
\sin\left(\frac{2\pi t}{p}+\phi^{\prime}\right)\\
0\\
\cos\left(\frac{2\pi t}{p}+\phi^{\prime}\right)\end{array}\right)+\delta a^{\prime i}\left(t\right)\label{orbit}\end{equation}
 where $\phi^{\prime}$ represents a constant phase term, and where
$\delta a^{\prime i}$ is small. Rotating from the primed system first
about the \emph{y}-axis, then the \emph{x}-axis, then the \emph{z}-axis,
we have
\begin{equation}
a^{i}\left(t\right)=l\left(\begin{array}{c}
\cos\Psi\sin\left(\frac{2\pi t}{p}+\phi\right)+\sin\Psi\sin\Theta\cos\left(\frac{2\pi t}{p}+\phi\right)\\
-\sin\Psi\sin\left(\frac{2\pi t}{p}+\phi\right)+\cos\Psi\sin\Theta\cos\left(\frac{2\pi t}{p}+\phi\right)\\
\cos\Theta\cos\left(\frac{2\pi t}{p}+\phi\right)\end{array}\right)+\delta a^{i}\left(t\right)\label{rotatedorbit}
\end{equation}
where $\phi$ subsumes rotation about the \emph{y}-axis with $\phi^{\prime}$
and where $\Theta$ and $\Psi$ are the angles of rotation of the
plane of motion away from the \emph{xz}-plane about the \emph{x}-
and \emph{z}-axes respectively.

\subsubsection{Solution to the geodesic equation}

The theory of the effects of small perturbations to the metric on
light propagation in the weak-field limit is already developed \cite{key-14,key-15}.
However, since the effects of a quadrupolar perturbation fall off
as $d^{3}$, it is desirable to expand the theory to be applicable
to regions of stronger fields. We note in particular that for a closely-orbiting
compact binary system such as a SMBHB, then \emph{m} and \emph{l} will be of similar
magnitude; therefore, we extend the first-order theory of light deflection
to order $\mathcal{O}\left(m/d\right)^{3}$.

First, note that all terms in (\ref{eq:smetric}) are $\mathcal{O}\left(m/r\right)$
or higher and that all terms in (\ref{eq:hmetric}) are of $\mathcal{O}\left(ml^{2}/r^{3}\right)\leq\mathcal{O}\left(m^{3}/r^{3}\right)$.
Let $\mathcal{O}\left(m^{3}/r^{3}\right)$ be small such that all
higher orders are negligible. Then, suppressing negligible terms,
\begin{equation}
\Gamma_{\beta\gamma}^{\alpha}=-\frac{1}{2}\left[\begin{array}{c}
\left(\eta^{\alpha\delta}+s^{\alpha\delta}\right)\left(s_{\beta\delta,\gamma}+s_{\gamma\delta,\beta}-s_{\beta\gamma,\delta}\right)\\
+\left(\eta^{\alpha\delta}\right)\left(h_{\beta\delta,\gamma}+h_{\gamma\delta,\beta}-h_{\beta\gamma,\delta}\right)\end{array}
\right] \, .\label{eq:Christoffel}\end{equation}
Let the Christoffel symbol associated with the Schwarzschild metric
$\Gamma_{\beta\gamma}^{\alpha\textrm{(S)}}\equiv-\left(1/2\right)\left(\eta^{\alpha\delta}+s^{\alpha\delta}\right)\left(s_{\beta\delta,\gamma}+s_{\gamma\delta,\beta}-s_{\beta\gamma,\delta}\right)$
and the remaining part resulting from the perturbation $\Gamma_{\beta\gamma}^{\alpha\textrm{(h)}}\equiv-\left(1/2\right)\left(\eta^{\alpha\delta}\right)\left(h_{\beta\delta,\gamma}+h_{\gamma\delta,\beta}-h_{\beta\gamma,\delta}\right)$.
Then (\ref{geodesic}) becomes
\begin{equation}
\dot{\kappa}^{\alpha}+\dot{\Xi}^{\alpha}+\left(\Gamma_{\beta\gamma}^{\alpha\textrm{(S)}}+\Gamma_{\beta\gamma}^{\alpha\textrm{(h)}}\right)\left(k^{\beta}+\kappa^{\beta}+\Xi^{\beta}\right)\left(k^{\gamma}+\kappa^{\gamma}+\Xi^{\gamma}\right)=0\label{eq:full geodesic} \, . 
\end{equation}
Since all $\Gamma_{\beta\gamma}^{\alpha\textrm{(S)}}$ and all components
of $\kappa^{\alpha}$ must be at least of $\mathcal{O}\left(m/r\right)$
or zero, (\ref{eq:full geodesic}) expands, again suppressing negligible
terms, to
\begin{equation}
\dot{\kappa}^{\alpha}+\dot{\Xi}^{\alpha}+\Gamma_{\beta\gamma}^{\alpha\textrm{(S)}}\left(k^{\beta}+\kappa^{\beta}\right)\left(k^{\gamma}+\kappa^{\gamma}\right)+\Gamma_{\beta\gamma}^{\alpha\textrm{(h)}}k^{\beta}k^{\gamma}=0\label{eq:Geodesic suppressed}\, .
\end{equation}
Since 
\begin{equation}
\dot{\kappa}^{\alpha}+\Gamma_{\beta\gamma}^{\alpha\textrm{(S)}}\left(k^{\beta}+\kappa^{\beta}\right)\left(k^{\gamma}+\kappa^{\gamma}\right)=0 \, ,\label{Schwarzschild geodesic}\end{equation}
we conclude
\begin{equation}
\dot{\Xi}^{\alpha}+\Gamma_{\beta\gamma}^{\alpha\textrm{(h)}}k^{\beta}k^{\gamma}=0\label{eq:Xi geodesic}
\end{equation}
which is exactly the result for the weak-field approximation \cite{key-14,key-15}.

\begin{figure}[p]
\includegraphics[width=1\textwidth]{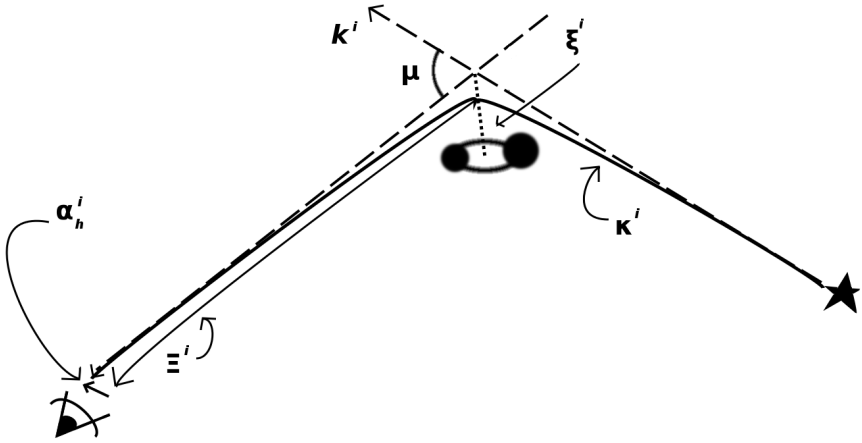}
\caption{\label{cap:Figure 2}A light ray produced by the emitter initially
follows trajectory $k^{i}$, which has its closest approach to the
origin of the coordinate system at $\xi^{i}$. In a pure Schwarzschild
space, the light ray follows trajectory $k^{i}+\kappa^{i}\left(\lambda\right)$
and is deflected about the apse of its trajectory by angle $\mu$;
in a perturbed Schwarzschild space, it follows trajectory $k^{i}+\kappa^{i}\left(\lambda\right)+\Xi^{i}\left(\lambda\right)$
and the observer records an additional deflection of $\alpha_{h}^{i}$.}
\end{figure}

Plugging (\ref{Schwarzschild geodesic}) and (\ref{eq:Xi geodesic})
into (\ref{alpha}) and choosing $\tau$ as our affine parameter,
we can define the Schwarzschild ($\alpha_{M}^{i}$) and non-Schwarzschild($\alpha_{h}^{i}$) parts of the
deflection angle{[}\noun{Figure} \ref{cap:Figure 2}{]}:
\begin{equation}
\alpha_{M}^{i}\left(\xi^{i}\right)\equiv P_{j}^{i}\kappa^{j}\label{eq:Schwarzshild alpha}
\end{equation}
and
\begin{equation}
\alpha_{h}^{i}\left(t,\xi^{i}\right)\equiv P_{j}^{i}\Xi^{j}=-\frac{1}{2}P^{ij}\int_{-\infty}^{\infty}\left(h_{\beta\delta,j}+h_{j\delta,\beta}-h_{\beta j,\delta}\right)k^{\beta}k^{\gamma}d\tau.\label{perturbed alpha}
\end{equation}
The monopole term $\alpha_{M}^{i}\left(\xi^{i}\right)$ of the deflection
produced by the core is static and unique, regardless of changes of
configuration within the core \cite{key-13,key-5}. We can use the
general, exact solution for $\kappa^{\alpha}$ provided by Darwin \cite{key-29}:

For purposes of this derivation only, choose spherical coordinates. By the symmetry of the monopole term,
this part of the trajectory of the light ray must lie in a plane,
so we can choose the coordinate $\theta$ as an affine parameter and
the coordinate $\phi$ as constant. Then we obtain an equation of
motion
\begin{equation}
-\frac{r-2m}{r}\left(\frac{dt}{d\theta}\right)^{2}+\frac{r}{r-2m}\left(\frac{dr}{d\theta}\right)^{2}+r^{2}=0.\label{eq:SchwarzschildEOM}\end{equation}
Identifying the impact parameter with a conserved quantity in the
system\\ $\left(r^{3}/\left(r-2m\right)\right)\left(dt/d\theta\right)=d$ and substituting
in, have three solutions; we discard the two where the light ray never
reaches a distant observer and take the remaining one,
\begin{equation}
\frac{1}{r\left(\theta\right)}=-\frac{Q-P+2m}{4mP}+\frac{Q-P+6m}{4mP}\textrm{sn}^{2}\zeta\left(\theta\right),\label{eq:u(theta)}\end{equation}
where the constant \emph{Q} is defined by $Q^{2}\equiv\left(P-2m\right)\left(P+6m\right)$,
the periastron and impact parameter are related by $d^{2}\equiv P^{3}/\left(P-2m\right)$
and\\ $\zeta\left(\theta\right)\equiv\sqrt{\left(Q/P\right)}\left(\theta+\theta_{0}\right)$,
and $\textrm{sn}{}\zeta$ is the Jacobi elliptic sn function \cite[16.1.5]{key-49}.
In the limit of $P\gg m$, inverting this relationship and taking
its asymptotic limits at large \emph{r} leads to the well-known relationship
\begin{equation}
\alpha_{M,\textrm{weak field}}^{i}\left(\xi^{i}\right)=\frac{4m}{d}n^{i}.\label{classicmonopole}\end{equation}
As $P\rightarrow3m$, however, the deflection becomes \cite{key-44}
\begin{equation}
\mu\left(\xi^{i}\right)=\ln\frac{m}{d}+\ln\left[648\left(7\sqrt{3}-12\right)\right]-\pi\approx\ln\frac{m}{d}+1.248\label{logmonopole}\end{equation}
where $\mu$ is the angle of deflection about the apse of the trajectory,
rather than the deflection seen by a distant observer; the angles
involved are no longer necessarily small so we cannot approximate
$\alpha_{M}=\mu$. In the case of an impact parameter comparable to
$3m$, it is no longer useful to consider the monopolar displacement
in and of itself as small differences in impact parameter cause great
changes in deflection angle, and multiple images of a source may be
detectable, some of which may result from geodesics which travel several
times around the deflector. Our consideration therefore must focus
not on the static deflection but on time-dependent deflections arising
from higher multipole moments of the deflector.

Kopeikin and Mashhoon \cite{key-41} develop the effect of the rotation
of a system on that system's deflection of light, in the weak field
approximation. Investgation of this effect is useful for the system
as described in that every practical case of an astronomical binary
will display orbital motion. However, the theory developed by Kopeikin
and Mashhoon is only sometimes compatible with the strong-field approximation
presented herein.

The integration of (\ref{eq:hqmetric}) is trivial. Let Let $\alpha_{S}^{i}\left(\xi^{i}\right)$
be that part of $\alpha_{h}^{i}$ determined by $h_{\mu\nu}^{S}$.
when the deflector is stationary relative to the observer, the resulting
contribution is given by
\begin{eqnarray}
\alpha_{S}^{i}\left(\xi^{i}\right) & =\frac{2}{d^{2}} & \left(2S^{j}k^{k}n^{l}\epsilon_{jkl}n^{i}+k^{j}S^{k}\epsilon^{i}{}_{jk}\right).\label{alphaspin}\end{eqnarray}

Calculating $S^{i}$ with (\ref{rotatedorbit}) for the case of a
binary whose components are in almost-circular orbits,
\begin{equation}
S^{i}=-m\frac{q}{1+q}\frac{2\pi l^{2}}{p}\left(\begin{array}{c}
\frac{q}{\left(1+q\right)^{2}-q^{2}\left(2\pi l/p\right)^{2}}\\
+\frac{1}{\left(1+q\right)^{2}-\left(2\pi l/p\right)^{2}}\end{array}
\right)\left(\begin{array}{c}
\sin\Psi\cos\Theta,\\
\cos\Psi\cos\Theta,\\
-\sin\Theta\end{array}\right).\label{spinvectosr}\end{equation}
We must emphasize that (\ref{alphaspin}) is compatible with the
$\mathcal{O}\left(m^{3}/r^{3}\right)$ generalization above only when
$\mathcal{O}\left(ml^{2}/d^{2}p\right)\geq\mathcal{O}\left(m^{3}/d^{3}\right)$;
in particular, when $l\approx p$ the system's motion is no longer
slow. We draw attention to this contribution to emphasize the difficulty
in asssociating an image with a particular source and to underscore
the utility of time-dependent deflection versus time-independent deflection
in parameterizing a system.

Let $\alpha_{Q}^{i}\left(t,\xi^{i}\right)$ be that part of $\alpha_{h}^{i}$
determined by $h_{\mu\nu}^{Q}$. $\alpha_{Q}^{i}$ is determined by
plugging (\ref{eq:hmetric}) into (\ref{perturbed alpha}); while
 \cite{key-15} uses the method of Fourier transforms, the form of
(\ref{eq:hmetric}) allows direct integration of a Fourier series
decomposition as well; either way, the result is the following deflection \cite{key-14,key-15}%
\footnote{The symbol $\epsilon_{ijk}$ represents the Levi-Civita permutation
symbol defined such that $\epsilon_{123}=1$.%
}:
\begin{equation}
\alpha_{Q}^{i}\left(t,\xi^{i}\right)=\frac{12}{d^{3}}\frac{mq}{\left(1+q\right)^{2}}\left[\left(a_{2}^{2}\left(s\right)-a_{1}^{2}\left(s\right)\right)n^{i}-a_{1}\left(s\right)a_{2}\left(s\right)\epsilon^{i}{}_{jk}k^{j}n^{k}\right]\label{quaddeflectionexplicit}\end{equation}
for which we reiterate the following properties: firstly, in contrast
to the monopolar case where $\alpha_{M}^{i}$ always points along
$\xi^{i}$, the quadrupolar deflection has a contribution parallel
to$\xi^{i}$, $\alpha_{Q\parallel}$, and also a contribution perpendicular
to $\xi^{i}$, $\alpha_{Q\perp}$, which vanishes only in the case
that a component of the projected quadrupole moment vanishes, that
is, only if the axis of rotation of the deflector is perpendicular
to our line of sight; and secondly, the deflection depends only on
the configuration of the deflector at the time of the light ray's
closest approach to the center of mass, $t=t^{*}$. 

In the case of almost-circular motion, inserting (\ref{rotatedorbit})
into (\ref{quaddeflectionexplicit}) leads to

\begin{eqnarray}
\nonumber\lefteqn{\alpha_{Q\parallel}^{i}\left(t,\xi^{i}\right)=} \\
& & \frac{12l^{2}}{d^{3}}\frac{mq}{\left(1+q\right)^{2}}\left\{ \begin{array}{c}
\frac{1}{2}\cos2\Psi\left[\begin{array}{c}
\left(1+\sin^{2}\Theta\right)\cos\left(\frac{4\pi s}{p}+2\phi\right)\\
+\sin^{2}\Theta-1\end{array}\right]\\
-\sin2\Psi\sin\Theta\sin\left(\frac{4\pi s}{p}+2\phi\right)\end{array}\right\} n^{i}
\label{circlimit}
\end{eqnarray}
and
\begin{eqnarray}
\nonumber\lefteqn{\alpha_{Q\perp}^{i}\left(t,\xi^{i}\right)=} \\
& & -\frac{6l^{2}}{d^{3}}\frac{mq}{\left(1+q\right)^{2}}\left\{ \begin{array}{c}
\frac{1}{2}\sin2\Psi\left[\begin{array}{c}
\left(-1+\sin^{2}\Theta\right)\\
+\left(1+\sin^{2}\Theta\right)\cos\left(\frac{4\pi s}{p}+2\phi\right)\end{array}\right]\\
+\cos2\Psi\sin\Theta\sin\left(\frac{4\pi s}{p}+2\phi\right)\end{array}\right\} \epsilon^{i}{}_{jk}k^{j}n^{k}.
\label{circlimit2}
\end{eqnarray}

The relationships (\ref{circlimit}) and (\ref{circlimit2}) are original to this work and
have not previously appeared. From this relationship it is easy to
see that the time-dependent deflection of the emitter's image is periodic,
with a period half that of the orbit of the core's components.

The greatest time-dependent deflection is observed when the emitter
lies on the line of the semimajor axis of the apparent motion; when
$q=1$; and when the plane of the system lies perpendicular to the
plane of observation. In this case, (\ref{circlimit}) reduces to
\begin{equation}
\alpha_{Q}\left(t,d\right)\leq\frac{3l^{2}}{2d^{3}}m\left[\cos\left(\frac{4\pi s}{p}+2\phi\right)-1\right]\label{idealized}\end{equation}
so the total quadrupolar deflection seen over one half-period of the
deflector's motion is
\begin{equation}
\Delta\alpha_{Q}\left(d\right)\leq-\frac{3l^{2}}{d^{3}}m.\label{idealizedtotal}\end{equation}
Compared to the monopole deflection in the case of a large impact
parameter,
\begin{equation}
\left|\frac{\Delta\alpha_{Q}}{\alpha_{M}}\right|\leq\frac{3l^{2}}{4d^{2}}.\label{compareMQ}\end{equation}

\subsubsection{Other contributions to the deflection angle}

If the path of the light ray after its closest approach to the deflector
but far from the deflector is nearly occulted (for example the Sun
or another star), then deflection from this intermediate deflector,
$\alpha_{\textrm{intermediate}}^{i}\left(\xi_{n,\textrm{int}}^{i}\right)$,
must be taken into account as well. Where$\xi_{n,\textrm{int}}^{i}$
refer to the vector impact parameters of light relative to these intermediate
deflectors, $d_{n}\equiv\left|\xi_{n,\textrm{int}}^{i}\right|$, and
$m_{n}$ are the masses of these deflectors, and where $m_{n}/d_{n}$
is small for all \emph{n},
\begin{equation}
\alpha_{\textrm{intermediate}}^{i}\left(\xi_{n}^{i}\right)=-\frac{4m_{n}}{d_{n}}\frac{\xi_{n,\textrm{int}}^{i}}{d_{n}}.\label{intermediate}\end{equation}

In linear approximation and in the harmonic gauge, the various deflections
can be superposed linearly. The total deflection of the light ray
from our source, therefore, is given by
\begin{equation}
\alpha^{i}\left(t,\xi^{i}\right)=\alpha_{Q}^{i}\left(t,\xi^{i}\right)+\alpha_{M}^{i}\left(\xi^{i}\right)+\alpha_{S}^{i}\left(\xi^{i}\right)+\alpha_{\textrm{intermediate}}^{i}\left(\xi_{n,\textrm{int}}^{i}\right).\label{totaldeflection}\end{equation}

\subsection{Application to 3C66B}

3C66B, also known as 0220+43, is a radio galaxy \cite{key-20} with
$z=0.0215$ \cite{key-21}, approximately 91 Mpc distant from the Milky
Way%
\footnote{We use a value of 71 km/s/Mpc for the Hubble constant $H_{0}$ for
all distance calculations. \cite{key-22}%
}. 3C66B exhibits jets emerging from its core, making it a good candidate
for the location of a SMBHB \cite{key-23}.

\subsubsection{Parameters of the system}

Sudou \emph{et al}. \cite{key-24} give estimates of 3C66B's parameters by 
direct radio observation of its core, including an upper limit on $m$,
a period, and an orientation of the core's motion. Sudou also reports
a limit on the minimum impact parameter available for determining
the parameters of the system using a first-order approximation theory,
corresponding to the limit of optical transparency at VLBI's higher
operating frequency. The parameters Sudou gives are:
\begin{equation}
\begin{array}{l}
m\leq4.4\left(1+q\right)^{2}\times10^{10}\textrm{ solar mass}=6.5\left(1+q\right)^{2}\times10^{15}\textrm{ cm},\\
l\leq5.1\left(1+q\right)\times10^{16}\textrm{ cm},\\
P\geq23\mu\textrm{ as}=3.1\times10^{16}\textrm{ cm},\\
d\geq3.7\times10^{16}\textrm{ cm},\\
p=1.05\pm0.03\textrm{ years},\\
\Theta=15^{\circ}\pm7^{\circ},\end{array}\label{sudou}
\end{equation}
where \emph{P} is constrained by the limit of the core's opacity in
the radio spectrum and $\Theta$ is derived from the apparent eccentricities
of the elliptical boundaries of radio opacity. From $l$ and $P$
we can furthermore conclude that in the case of maximized $l$, $q\leq0.20$.

\subsubsection{Estimates for distant emitters}

Although highly eccentric motion in 3C66B is not ruled out \cite{key-26},
the age of the presumed binary is great enough to have circularized
the orbit through gravitational radiation under most conditions \cite{key-28}.
We present the case of circular motion as an upper limit on the time-dependent
deflection angle, noting that if all other parameters are constant
then in the case of eccentric motion any time-dependent separation
of the masses must have \emph{l} as an upper bound. 

Using the maximal figure for mass and the minimal figure for impact
parameter in (\ref{sudou}) and applying (\ref{logmonopole}), the
ratio $m/l=0.30$, placing our proposed system in the regime of strong
deflection. We find a monopolar deflection of:
\begin{equation}
\mu=\ln\left(\frac{6.5\times10^{15}\textrm{ cm}}{3.7\times10^{16}\textrm{ cm}}\left(1.44\right)\right)+1.248=0.13\textrm{ radian}=7.2^{\circ}.\label{eq:3c66b mu}\end{equation}

The components of the system as proposed by Sudou have $2\pi l/p\leq0.39$.
Therefore it is not reasonable to apply (\ref{alphaspin}) to 3C66B
in the regime where deflection from the quadrupole moment will be
detectable.

Deflected images lying along the major axis of the core with the system
as constrained in (\ref{sudou}) will have time-dependent deflections
in the following amounts:
\begin{equation}
\begin{array}{c}
\Delta\alpha_{Q\parallel}\left(d\right)\leq\frac{12l^{2}}{d^{3}}\frac{mq}{\left(1+q\right)^{2}}\left(1.07\right),\\
\Delta\alpha_{Q\parallel}\left(d\right)\leq
\begin{array}{c}
\frac{12\left(5.1\times10^{16}\textrm{ cm}\right)^{2}\left(1.2\right)^{2}}{\left(3.7\times10^{16}\textrm{ cm}\right)^{3}}\frac{\left(6.5\times10^{15}\textrm{ cm}\right)\left(1.2\right)^{2}\left(0.2\right)}{\left(1.2\right)^{2}}\left(1.07\right)\\
=5.8\times10^{-5}\textrm{ arcsecond}\end{array};\end{array}\label{3c66bparallel}\end{equation}

\begin{equation}
\begin{array}{c}
\Delta\alpha_{Q\perp}\left(d\right)\leq\frac{12l^{2}}{d^{3}}\frac{mq}{\left(1+q\right)^{2}}\left(0.26\right),\\
\Delta\alpha_{Q\perp}\left(d\right)\leq
\begin{array}{c}
\frac{12\left(5.1\times10^{16}\textrm{ cm}\right)^{2}\left(1.2\right)^{2}}{\left(3.7\times10^{16}\textrm{ cm}\right)^{3}}\frac{\left(6.5\times10^{15}\textrm{ cm}\right)\left(1.2\right)^{2}\left(0.2\right)}{\left(1.2\right)^{2}}\left(0.26\right)\\
=1.4\times10^{-5}\textrm{ arcsecond}\end{array}\end{array}\label{3c66bperp}\end{equation}
with a period of $p/2=0.53\pm0.02\textrm{ years}$for each component
of the deflection.

\section{Observational techniques}

\subsection{Interferometry}

Electromagnetic interferometry provides the best currently-available
techniques for high-resolution astronomy. The use of space-based interferometry
and improvements in equipment allowing for higher frequencies of observation
continue to steadily improve resolution capabilities. The current
most powerful technique available is VLBI, which Sudou \emph{et al.}
used to determine the motion in the core of 3C66B \cite{key-24}.

VLBA, the Very Long Baseline Array, is an array of ten radio telescopes \cite{key-32}
operating in wavelengths as short as 3mm operating as a single large
interferometer. The current best available resolution is\\ $1.7\times10^{-5}\textrm{ arcsecond}$ \cite{key-33},
making VLBA currently capable of constraining the parameters of 3C66B
further through direct observation as well as the Jenet pulsar timing
experiment described below accomplishes indirectly. The launch of
the space-based ASTRO-G satellite \cite{key-34} will extend the resolution
capabilities further.

The SIM PlanetQuest mission (formerly Space Interferometry Mission),
currently scheduled for launch in 2015 \cite{key-35}, is expected
to have a resolution capability of $4\times10^{-6}\textrm{ arcsecond}$ \cite{key-36}.
SIM will operate in the optical band and quasar observation is part
of the planned mission.

Farther into the future, the MAXIM (Micro-Arcsecond X-ray Interferometry
Mission) satellite array currently in development \cite{key-38} is
expected to give resolutions on the order of $10^{-7}\textrm{ arcsecond}$
in the x-ray band, and is explicitly designed with the observation
of black holes in mind.

\subsection{Pulsar timing}

Jenet \emph{et al.} \cite{key-8} examined the period of the pulsar
PSR B1855+09 for changes in its period over several years, motivated
by the idea that as gravitational waves generated by the core of 3C66B
pass through the pulsar then the pulsar's signal should be modulated
with a period related to the period of the proposed 3C66B SMBHB. The
distance between the Solar System and the pulsar furthermore give
the advantage that the signals observed modulating the pulsar are
some 4000 years older than the motion observed in the 3C66B core, providing some information about the evolution of the system.
However, Jenet's experiment produced a null result.

The experiment's analysis involved examining the frequency space of
the pulsar's signal for components in a range from $\left(1/27.8\right)\textrm{ yr}^{-1}$
to $19.5\textrm{ yr}^{-1}$, then subtracting out the one-year and
six-month components resulting from geodetic effects. The results
are described as showing no signal distinguishable from noise other
than the already-known main oscillation frequencies of the pulsar.
Therefore the magnitude of gravitational waves generated by 3C66B,
and consequently the parameters of its core, can be further constrained.

Jenet \emph{et al.} assert that a system with\\ $m\left(q/\left(1+q\right)^{2}\right)^{3/5}\geq0.7\times10^{10}\textrm{ solar mass}$
can be ruled out by the observed null result in the change in pulsar
periods over seven years; this corresponds in the optimal case of
$q=0.2$ to a system with $m=2.3\times10^{10}\textrm{ solar mass}=3.4\times10^{15}\textrm{ cm}$.
For a system under these new constraints, we estimate optimal peak
deflections:
\begin{equation}
\begin{array}{c}
\Delta\alpha_{Q\parallel}\left(d\right)\leq
\begin{array}{c}
\frac{12\left(5.1\times10^{16}\textrm{ cm}\right)^{2}\left(1.2\right)^{2}}{\left(3.1\times10^{16}\textrm{ cm}\right)^{3}}\left(3.4\times10^{15}\textrm{ cm}\right)\frac{\left(0.2\right)}{\left(1.2\right)^{2}}\left(1.07\right)\\
=2.1\times10^{-5}\textrm{ arcsecond}\end{array};\\
\Delta\alpha_{Q\perp}\left(d\right)\leq
\begin{array}{c}
\frac{12\left(5.1\times10^{16}\textrm{ cm}\right)^{2}\left(1.2\right)^{2}}{\left(3.1\times10^{16}\textrm{ cm}\right)^{3}}\left(3.4\times10^{15}\textrm{ cm}\right)\frac{\left(0.2\right)}{\left(1.2\right)^{2}}\left(0.26\right)\\
=5.0\times10^{-6}\textrm{ arcsecond}\end{array}\end{array}\label{jenetestimates}\end{equation}
which remains within the detection limit of VLBA as currently configured.

\section{Conclusions}

A theory of light deflection by time-dependent distributions of matter
has been presented for metrics which are perturbations of the Schwarzschild
metric, accounting for deflection resulting from time-independent
and time-dependent terms in the metric. To order $m^{3}/r^{3}$, deflections
originating from the quadrupole moment of the mass distribution and,
with some constraints, the dipole moment of the system's spin can
be linearly superposed on the system as if in a weak-field approximation.
The theory can be practically evaluated for and applied to a model
of the core of an active galaxy, but the theory of light deflection
from the spin of the deflector needs further development for applicability
in the regime of strong deflection.

The examination of time-dependent light deflection is a feasible technique
for the evaluation of proposed SMBHB systems, under idealized circumstances.
In the event that a suitable emitter exists, examination of light
deflection can be used to constrain the parameters of the proposed
SMBHB in the core of 3C66B. We emphasize that while the existence
of an identifiable suitable emitter in the case of 3C66B is unlikely,
the theory can be applied equally well to any other SMBHB candidate,
any of which may have a suitable source; in particular, active galaxies
displaying Einstein rings or other artifacts of strong gravitational
lensing, especially multiple images, should be examined. The theory
can be equally well applied to intragalactic objects.

The quadrupolar motion in the core of 3C66B can be examined and evaluated
by the observation of deflected images in the region of the sky near
the core of the galaxy, if found; the time-dependent part of the deflection
has a magnitude of up to 58 microarcseconds parallel to the impact
parameter of the emitter and up to 14 microarcseconds perpendicular
to the impact parameter.

For the case of 3C66B, for most emitters pulsar timing can constrain
the parameters of the deflecting system better than time-dependent
light deflection can. VLBA in its current configuration is capable
of constraining the parameters of the core of 3C66B under ideal circumstances.
Anticipated interferometers will have resolutions up to two orders
of magnitude greater and will be capable of evaluating the parameters
of the system more closely while examining it in a wide range of frequencies,
and may make the observation of time-dependent light deflection resulting
from motion in the core of 3C66B more practical.

\section*{Acknowledgments}

Thanks are given to Prof. Leonid P. Grishchuk of Cardiff University
for continued help and support. Prof. Sergei Kopeikin of University
of Missouri-Columbia, Dr. Simon Dye of Cardiff University, and Dr.
Naomi Ridge formerly of Harvard University provided helpful discussions.

\end{document}